\documentclass[aps,prb,twocolumn,amssymb,color,pdflatex]{revtex4}

\usepackage[pdftex]{graphicx}
\usepackage[pdftex]{graphics}
\usepackage{dcolumn}
\usepackage{bm}
\usepackage{amsmath}
\usepackage{subfig}
\usepackage{ifpdf}
\begin{document}

\newcommand{\bk}{{\bf k}}
\newcommand{\bp}{{\bf p}}
\newcommand{\bv}{{\bf v}}
\newcommand{\bq}{{\bf q}}
\newcommand{\tbq}{\tilde{\bf q}}
\newcommand{\tq}{\tilde{q}}
\newcommand{\bQ}{{\bf Q}}
\newcommand{\br}{{\bf r}}
\newcommand{\bR}{{\bf R}}
\newcommand{\bB}{{\bf B}}
\newcommand{\bA}{{\bf A}}
\newcommand{\ba}{{\bf a}}
\newcommand{\bE}{{\bf E}}
\newcommand{\bj}{{\bf j}}
\newcommand{\bK}{{\bf K}}
\newcommand{\cS}{{\cal S}}
\newcommand{\vd}{{v_\Delta}}
\newcommand{\tr}{{\rm Tr}}
\newcommand{\kslash}{\not\!k}
\newcommand{\qslash}{\not\!q}
\newcommand{\pslash}{\not\!p}
\newcommand{\rslash}{\not\!r}
\newcommand{\bs}{{\bar\sigma}}
\newcommand{\omt}{\tilde{\omega}}

\title{Inducing topological order in a honeycomb lattice}

\author{T. Pereg-Barnea$^1$ and G. Refael$^2$}
\affiliation{$^1$Department of Physics, McGill University, Montreal, QC, Canada}
\affiliation{$^2$Department of Physics,
California Institute of Technology, 1200 E. California Blvd, MC114-36,
Pasadena, CA 91125}

\date{\today}

\begin{abstract}
We explore the possibility of inducing a topological insulator phase
in a honeycomb lattice lacking spin-orbit interaction using a metallic
(or Fermi gas) environment.
The lattice and the metallic environment interact through a density-density
interaction without particle tunneling,  and integrating out the
metallic environment produces a honeycomb sheet with in-plane
oscillating long-ranged interactions.   We find the ground state of the interacting system
in a variational mean-field method and show that the Fermi wave
vector, $k_F$, of the metal determines which phase occurs in the
honeycomb lattice sheet. This is analogous to the Ruderman-Kittel-Kasuya-Yosida (RKKY) mechanism in which the metal's
$k_F$ determines the interaction profile as a function of the
distance. Tuning $k_F$ and the interaction strength may lead to a
variety of ordered phases, including a topological insulator and
anomalous quantum-hall states
with complex next-nearest-neighbor hopping, as in the Haldane and the
Kane-Mele model. We estimate the required range of
parameters needed for the topological state and find that the Fermi
vector of the metallic gate should be of the order of $3\pi/8a$ (with
$a$ being the graphene lattice constant).  The net coupling between
the layers, which includes screening in the metal, should be of the
order of the honeycomb lattice bandwidth. This configuration should be
most easily realized in a cold-atoms setting with two interacting Fermionic
species.

\end{abstract}
\maketitle

\section{Introduction}
In 2005 Kane and Mele\cite{KaneMele1,KaneMele2} proposed that
graphene, a one atom thick graphite can exhibit helical edge states
which are protected against weak perturbations by topology.  This idea
stems from the Haldane model\cite{Haldane} of a quantum Hall state
without magnetic field.  This special topological phase would have
emerged as the ground state of graphene if the intrinsic spin-orbit
coupling was large enough and in particular, larger than the Rashba
spin-orbit coupling.  Unfortunately, ab-initio calculations\cite{Min}
have found that this requirement is far from being fulfilled in
graphene and the desired intrinsic spin-orbit coupling is smaller than
0.001meV (about 0.01K).  Nevertheless, this direction has led the way
to a wealth of theoretical
predictions\cite{Bernevig,Guo1,Guo2,Ruegg,Zhang} of topological
states.  In parallel, experiments have shown that robust
helical edge states exist in three dimensions\cite{Hsieh,Xia} and in
two dimensional quantum wells\cite{Konig,Roth}.

Despite the fact that Graphene's intrinsic spin-orbit coupling may be
too weak to produce the desired topological insulator phase the hope
to achieve such a phase in the honeycomb lattice has not died.  In
this work we explore the possibility of engineering a
honeycomb-lattice-based or Graphene-based
topological insulator by employing an appropriate
environment. Our results are relevant either to a Graphene with a
metallic gate, or to a Fermi-gas residing in a honeycomb optical
lattice (see, e.g., Ref. \cite{Esslinger}), and interacting with
a second Fermi-gas of similar (two-dimensional) density that does not
couple to the optical lattice. Note that another path for producing
topological states in optical lattices was explored in
Ref. \onlinecite{Galitski-2d}.

An important ingredient for topological behavior is a bulk gap that
arises through avoided level crossing between two bands and a
resulting band inversion. A half filled honeycomb lattice (e.g., undoped graphene) has band crossing at the
two Dirac points \cite{Wallace,Esslinger} but without any band gap.  An opening
of a gap in the bulk is possible in a few ways, for example by
substrate effects\cite{Zhou} or a large lattice
distortion\cite{Bahat}.  However, not all gaps are alike.  In a two
valley model like graphene, non-trivial topology can only occur if the
gap function changes sign between the two valleys.  This sign change
can occur, for instance, due to intrinsic spin-orbit coupling.  On
the lattice, this coupling may be approximate by imaginary hopping
between next nearest neighbor sites  and this yields the Kane-Mele
model.  The sign of the imaginary hopping depends not only on the
direction of the hopping but also on whether the path from one site to
another contains a left or right turn.  In momentum space, the
imaginary hopping takes the form of a function that changes
sign between Dirac valleys, and between spin
polarizations.

Apart from intrinsic spin-orbit coupling, this type of directional
hopping term has been found to arise from an interaction between next
nearest neighbor sites\cite{Raghu,Weeks}.  It has also been found that
interactions may lead to a topological insulator in a decorated
honeycomb and Kagom\'e lattices\cite{Wen}.

The main obstacle for realizing an interaction-driven topological
insulator is the required interaction profile in real space.  In
a honeycomb lattice, for example, this amounts to an interaction which is
strongest on next-nearest-neighbor bonds.  When this requirement is
not fulfilled, other phases (density waves, lattice distortions, etc.)
may occur.  An on-site repulsion $U$ (arising due to Coulomb
interactions in electrons, or on-site scattering for cold atom), enhances the tendency
for magnetic ordering.  The next strongest interaction (for electrons)
is the nearest-neighbor repulsion, $V_1$, which tends to favor charge/spin density waves.
A second-nearest-neighbor interaction $V_2$, however, enhances
precisely the propensity to induce an imaginary hopping term leading
to the topological phase.  It is hard to find, however, lattice
systems where the second-nearest neighbor interaction trumps the
shorter term repulsions; therefore, the topological phase is
not likely to arise due to intrinsic interactions.  In addition, if $V_1$ and $V_2$ are of similar order, a
Kekul\'e bond order may occur\cite{Weeks}.  Extensive studies
on interacting graphene have shown a variety of competing
phases\cite{Herbut1,Herbut2,Herbut3}.
Below we discuss the effects of producing the longer range
interactions required for topological order using the Friedel oscillations and RKKY interactions
occurring in a metallic (Fermi-gas) environment. Our analysis considers
the possibility of forming
both topological phases as well as other competing phases.

\section{Tuning the interaction}
In this work we propose a setup in which a 'knob' can be turned to
control the relative ratio between the different interaction terms.
Our idea is motivated by the RKKY\cite{RKKY1,RKKY2,RKKY3} interaction
in metals.  The original RKKY model was written to describe how
impurity spins interact with each other over a long distance through a
medium of electronic states.  The interaction does not simply fall as
a power law of the distance between the spins like the usual Coulomb
interaction but instead oscillates.  The oscillations can be viewed as
Friedel oscillations in the metal caused by the impurity spins. The
interaction profile depends on the polarization of the metal and its
functional form is $-\cos(2k_Fr)/r^3$ in three dimensions and
$-\cos(2k_Fr)/r^2$ in two.  The Fermi gas background is considered as a simple parabolic band.

With this scenario in mind we propose to generate interactions in the
honeycomb lattice that are mediated by an additional Fermi gas or
metallic gate. An electronic realization of such a system consists of
a metallic layer which is put
above a honeycomb lattice such that the quasiparticles in it interact
with the metallic-gate quasiparticles without direct hopping between the
two systems. A more promising avenue, which our analysis more readily
addresses, is a cold-atoms realization, consisting of two
different species of Fermions, only one of which is trapped in an optical
honeycomb lattice, with a contact interaction between the two species.
The effect of the coupling between the honeycomb fermions and the free
gas can
be found by integrating out the metallic degrees of freedom.  This
results in the honeycomb fermions acquiring a long range interaction whose magnitude
depends on the polarization of the free Fermi gas. For simplicity, we
will assume that the fermions are electron-like, and refer to them as
electrons and fermions interchangeably below.

\section{Formalism}
\ifpdf
\begin{figure*}
  \centering
  \subfloat[$U=0$]{\includegraphics[width=0.32\textwidth]{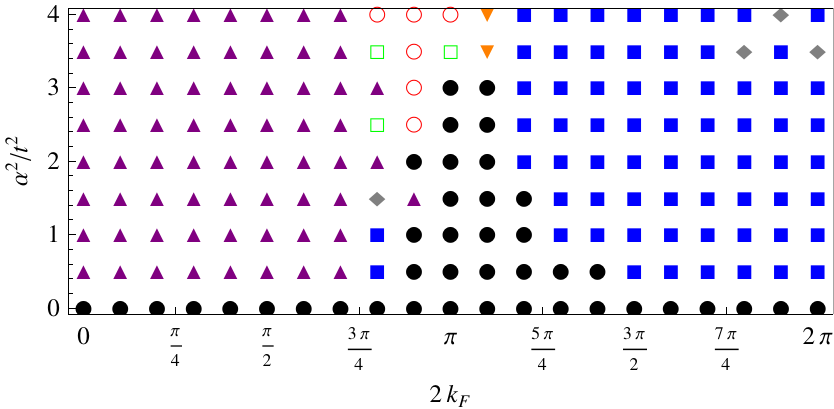}}
  \subfloat[$U=t$]{\includegraphics[width=0.32\textwidth]{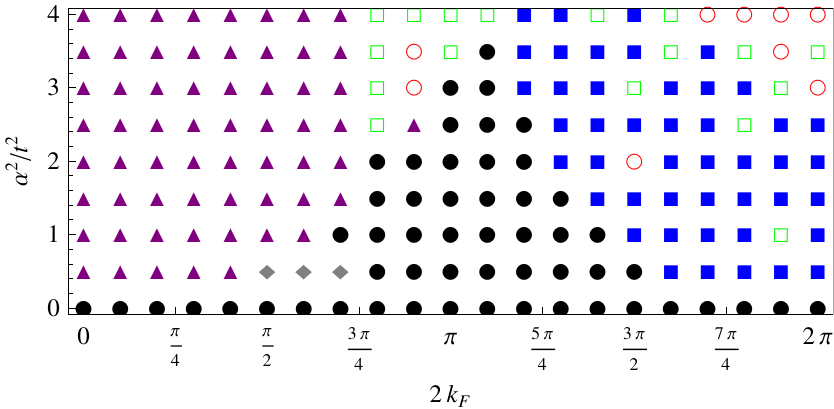}}
  \subfloat[$U=2t$]{\includegraphics[width=0.32\textwidth]{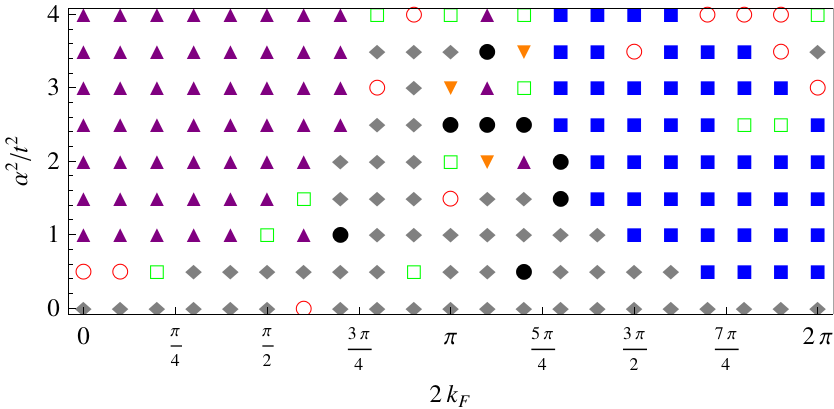}}
  \caption{Phase diagram for graphene with induced long ranged interaction of the form of Eq.~\ref{eq:RKKY}.  The $x$-axis is $2k_F$ of the metallic layer (measured in units of $1/a$ where $a=2.46\AA$ is the graphene lattice constant) which governs the relative strength of interaction in different distances and the $y$-axis is the overall interaction strength which is determined by the coupling $\alpha$.  The on-site interaction parameter $U$ was scanned.  Here we show (a) $U=0$, (b) $U=t$ and (c) $U=2t$ where $t$ is the hopping amplitude.  The different symbols (colors online) represent different phases with: circle (black) semi-metal, square (blue) charge density wave, diamond (gray) spin density wave, up-triangle (purple) superconductor, down-triangle (orange) Kekul\'e, open circle (red) anomalous Hall, open square (green) anomalous spin Hall.}
\label{fig:spinfull}
\end{figure*}
\fi

Our starting point is electrons hopping on a honeycomb lattice and interacting through an on-site density-density interaction with fermions in a metallic layer:
\begin{eqnarray}\label{eq:int}
{\cal H} = -t\sum_{\langle ij \rangle}\sum_\sigma c_{i \sigma}^\dagger c_{j\sigma} + \sum_{\bk \sigma} \epsilon_\bk d_{\bk \sigma}^\dagger d_{\bk \sigma} +\alpha\sum_i n_i^c n_i^d
\end{eqnarray}
where $c_{i\sigma}^\dagger$ creates an electron with spin $\sigma$ on
the honeycomb lattice site $i$, $d_{\bk \sigma}^\dagger$ creates an
electron with spin $\sigma$ in a Bloch state $\bk$ of the metal and
$n_i^c$ and $n_i^d$ are the number operators at the position $i$ of
the honeycomb lattice and metal electrons respectively.  In the metal this
should be understood as a coarse grained density integrated over a
small area, $A_{eff}$ around the position $i$.  This area should be
estimated by the screening in the metal, or in the cold-atom case, by
the probability density of the two species. We set it to the lattice constant $a$ in order to simplify the notation.  A larger area may enhance the interaction but will also make the approximate interaction term in Eq.~\ref{eq:int} less accurate.  We estimate the value of $\alpha^2$ in the discussion section and in the appendix.

In the path integral language the action is given by a time integration over the Lagrangian which is derived from the above Hamiltonian and the integration over the metallic Fermion operators $d_k$ is possible due to their quadratic form.  The result is an interacting theory of quasiparticles on the graphene sheet with their action given by:
\begin{eqnarray}
\int d\tau \left[ t\sum_{\langle ij \rangle}\sum_\sigma c_{i \sigma}^\dagger c_{j\sigma} + \alpha^2 \int d\tau' \sum_{ij}\Pi_{ij}(\tau-\tau')n_i^c n_j^c \right]
\end{eqnarray}
where the integration is over the imaginary time variables $\tau$ and $\tau'$ and $\Pi_{ij}(\tau-\tau')$ is the polarization operator of the metal.  The interaction above is long ranged and dynamic.  Its functional form in momentum and Matsubara frequency space is given by the Matsubara sum of the polarization bubble of the Fermi sea:
\begin{eqnarray}
\Pi(i\Omega,\bq) = \int d^2k{n_F(\epsilon_\bk)-n_F(\epsilon_{\bk+\bq}) \over i\Omega - \epsilon_\bk + \epsilon_{\bk+\bq}}
\end{eqnarray}
where $n_F$ is the Fermi-Dirac occupation number, which includes the chemical potential.  This polarization operator leads to a long ranged, time dependent interaction in the honeycomb lattice.  In the current work, for simplicity, we explore the limit of static interaction.  The static limit is valid if the velocity in which information travels in the honeycomb lattice is much slower than the typical velocity in the metal.  The comparison can be made in terms of energy.  In the graphene sheet we take the energy of the gap that opens as a result of the topological order and in the metal it is the Fermi energy.  We found that there is indeed a range of parameters where the developed gap is small compared to the metal's Fermi energy.

\section{Method and results}
The static limit of the polarization bubble is the famous Lindhard function in two dimensions\cite{Simon,Giuliani}.  It is given by:
\begin{eqnarray}\label{eq:RKKY}
\Pi(\Omega\to 0,r) = -{a^2\over t}{\cos(2 k_F r) \over r^2}
\end{eqnarray}
where $r$ is the distance between the interacting sites, $k_F$ is the Fermi wavevector of the metal, the $a^2$ factor represents the effective interaction area $A_{eff}$ and $1/t$ is the estimated metallic density of states at the Fermi level.  In order to make a crude estimate of the effective interaction strength in the layer we assume that the interaction between the electrons on the graphene sheet and the metal electrons is $e^2/d$ where the distance $d$ is of the order of $10\AA$.  This gives a coupling $\alpha$ which is of the order of the bandwidth $t$ (2-3eV).  However, when screening is taken into account (see the appendix) the bare coupling is reduced by a factor of $\exp(-4k_F d)$ and therefore the required distance is much smaller.  This poses a great challenge since at small distances tunneling may occur.  We present results (Fig.~\ref{fig:spinfull}) for this order of interactions where $(\alpha/t)^2$ is scanned from 0 to 4.

\ifpdf
\begin{figure}[h]
\includegraphics[width=\columnwidth]{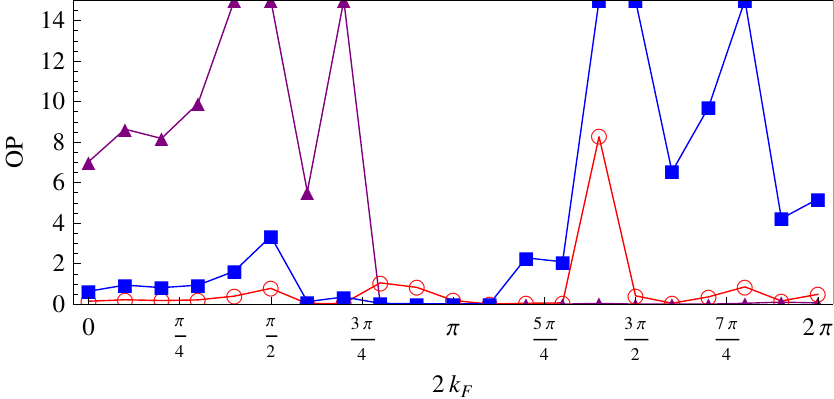}
\caption{An example of order parameter magnitudes as a function of the metal's Fermi wavelength (measured in units of $1/a$ where $a=2.4\AA$ is the graphene lattice constant).  The data is given for a constant $U=0$ and $V_0=4t$.  The curves (color online) represent the order parameter magnitude in units of the bare hopping $t$ with: (blue) square - charge density wave, (red) circle - anomalous Hall and (purple) triangle - superconductivity.}\label{fig:OPs}
\end{figure}
\fi

The standard mean field decomposition leads to self-consistency equations that are usually solved by iterations.  The interaction terms are decomposed in the various channels.  For example, the next-nearest neighbor interaction has been decoupled by Raghu {\it et al.} into the desired second neighbor hopping interaction:
$V_2 c^\dagger_ic_i c^\dagger_jc_j \to V_2(\chi_{ij}c_j^\dagger c_i + h.c - |\chi_{ij}|^2)$ and the self-consistency equation reads $\chi_{ij}=\langle c_i^\dagger c_j\rangle$.  This decomposition is reasonable, however, one should take care to decouple {\it all} interaction terms in {\it all} channels.  When this is not done correctly the combined effect of multiple terms may result in errors and phases may be missed.  Given the drawbacks of the standard mean-field decoupling we adopt the variational mean field approach\cite{Chaikin}.  This is a systematic procedure which does not suffer from the above problems.  We generate a variational wavefunction that is a solution to a quadratic auxiliary Hamiltonian.  This generating Hamiltonian, ${\cal H}_{gen}$, contains the graphene nearest neighbor hopping and a variety of order parameters.  These include charge and spin density waves in which the two sublattices have different charge/spin density; the Kekul\'e distortion, which is a structural distortion known to occur in carbon nanotubes\cite{Chamon} and may have interesting topological excitations\cite{Hou}, the anomalous Hall or spin Hall state and superconductivity.  Superconductivity is included here since at certain Fermi vectors some interaction coefficients may be attractive.  The resulting superconducting state has an interesting momentum-dependent order parameter which resembles the graphene's band dispersion.  Near the valleys the superconducting order parameter has a $p_x\pm ip_y$ form with the sign changing between the two valleys\cite{Uchoa}.  Another interesting suggestion for this region is Kekul\'e superconductivity\cite{Roy}.

The generating Hamiltonian is a 12$\times12$ matrix for each momentum $\bk$ with the following structure.
\begin{eqnarray}\label{eq:genH}
{\cal H}_{gen}(\bk) = \begin{pmatrix} \hat h_\uparrow & \hat g \\ \hat g^\dagger & -\hat h_\downarrow \end{pmatrix} \nonumber \\
\hat h_\uparrow = \begin{pmatrix} H_{AA} & H_{AB} \\ H_{BA} & H_{BB}\end{pmatrix} \;\;\;\;
\hat g = \begin{pmatrix} G_{AA} & G_{AB} \\ G_{BA} & G_{BB}\end{pmatrix}
\end{eqnarray}
where the $2\times2$ structure of ${\cal H}_{gen}$ is the Nambu space with $\hat h$ containing the usual particle-hole terms and $\hat g$ containing the pairing amplitude.  The matrices $\hat h$ and $\hat g$ are further split into the $A$ and $B$ sublattices and due to the Kekul\'e distortion (which enlarges the unit cell to include three atoms of each sublattice) the matrices $H_{nm}$ and $G_{nm}$ of dimension 3.
\begin{eqnarray}
H_{AA} &=& \begin{pmatrix}
      \rho_\uparrow & S_{\bk\uparrow} & S^*_{\bk\uparrow} \\
      S^*_{\bk\uparrow} &\rho_\uparrow & S_{\bk\uparrow}  \\
      S_{\bk\uparrow} &S^*_{\bk\uparrow} &\rho_\uparrow
         \end{pmatrix} \nonumber \\
H_{AB} &=&
{\small \begin{pmatrix}
      t_3 & t_2e^{i\bk\cdot\ba_2}& t_1e^{i\bk\cdot\ba_3}\\
      t_2e^{i\bk\cdot\ba_3}& t_1 &  t_3e^{i\bk\cdot\ba_2}\\
      t_1e^{i\bk\cdot\ba_2}&  t_3 e^{i\bk\cdot\ba_3}& t_2 \\
         \end{pmatrix} }\nonumber \\
H_{BA} &=& H_{AB}^\dagger  \nonumber \\
G_{AA} &=& G_{BB} = 0  \nonumber \\
G_{AB} &=&\Delta \begin{pmatrix}
      1& e^{i\bk\cdot\ba_2}&e^{i\bk\cdot\ba_3} \\
             e^{i\bk\cdot\ba_3}&1&e^{i\bk\cdot\ba_2} \\
             e^{i\bk\cdot\ba_2}&e^{i\bk\cdot\ba_3}&1
         \end{pmatrix} \nonumber \\
G_{BA} &=& G_{AB}^\dagger.
\end{eqnarray}
Here $\rho_{\uparrow}$ and $\rho_{\downarrow}$ are the order parameters for density waves in the up and down spin respectively, $S_{\bk\uparrow/\downarrow} = i\lambda_{\uparrow/\downarrow} (e^{i\bk\cdot\ba_1}+e^{i\bk\cdot\ba_2}+e^{i\bk\cdot\ba_3})$ are the Fourier transform of the second nearest neighbor hopping with $\lambda_\uparrow$ and $\lambda_\downarrow$ being the order parameter for topological order in the up and down spin and the vectors $\ba_1,\ba_2$ and $\ba_3$ are (hexagonal) lattice vectors. ($\ba_1 = a\hat x$, $\ba_{2/3} = a({1\over 2}\hat x \pm {\sqrt{3}\over 2}\hat y)$).  Following Weeks and Franz\cite{Weeks} we parametrize the lattice distortion by $t_j = t+\delta t + \eta_j$ where $\delta t$ is a uniform shift and the Kekul\'e texture is given by $\eta_j = \eta \cos({2\pi\over 3}j+\phi)$ with $\eta$ being the order parameter and $\phi$ a free parameter which does not affect the size of the gap.  The down spin part of the generating Hamiltonian $\hat h_\downarrow$ is obtained from  $\hat h_\uparrow$ by replacing $\rho_\uparrow \to \rho_\downarrow$ and $\lambda_\uparrow \to \lambda_\downarrow$.

The trial wavefunction, which is the ground state of the Hamiltonian in Eq.~\ref{eq:genH} is used to calculate the full interacting variational energy.  This is done by calculating the expectation of the Hamiltonian:
\begin{eqnarray}
{\cal H} = -t\sum_{\langle ij \rangle}\sum_\sigma c_{i \sigma}^\dagger c_{j\sigma} +
 U\sum_i n_{i\uparrow} n_{i \downarrow}+\sum_{i \delta}V_\delta n_i n_{i+\delta}
\end{eqnarray}
where $U$ represents an effective on-site interaction and $V_\delta$ represents the interaction between two sites separated by the distance $\delta$ which is taken up to the seventh neighbor and is given by the coupling $\alpha^2$ times the RKKY interaction in Eq.~\ref{eq:RKKY}. The on-site interaction is estimated at 3.5eV (about 1.3t)\cite{Uestimate} but since the uncertainty is large, we choose to scan different values of $U$. The grid size we use is $64\times64$ unit cells (of six atoms each). In order to save computing time we calculate the correlation functions in momentum space, Fourier transform them and use the resulting real-space correlations to evaluate the interaction energy for any given distance.
The next step is to minimize the energy with respect to the order parameters and the minimum determines the mean-field ground state.  We use the standard downhill simplex method\cite{numerical} to do this minimization.  The results are shown in Figs.~\ref{fig:spinfull}-\ref{fig:OPs}.

\section{Summary and Discussion}

In this Paper we analyzed the effect of a metallic environment on
fermions residing in a honeycomb lattice. The main idea is that
the environment, through its Lindhard response, can induce long
range interactions in the honeycomb lattice, which are conducive to
the formation of topological phases, such as the anomalous and
spin-Hall states. Our analysis consisted of a variational study of the many possible
orders of a honeycomb lattice in the presence of an oscillating long-range interaction. To
test our analysis, and to make a connection with previously derived results we used an
interaction cut-off at the next nearest neighbor length and plot a
phase diagram for a spinless $V_1-V_2$ model.  Our results are similar
to those obtained by Weeks and Franz \cite{Weeks} when spinless
Fermions are considered.  It is interesting to note that the Kekul\'e
phase vanishes as soon as the spin and on-site interaction are introduced.

When we introduce the oscillating long-range component of the
interaction, we indeed find that topological phases, among other phases, may
be induced in a honeycomb lattice through interaction with a metallic
environment. In particular, topological order occurs in the case of strong coupling between the two species,
when the Fermi wave number of the metal is between $\pi/4a$ and $\pi/2a$
(where $a$ is the lattice constant of the honeycomb).  This is shown in Fig.~\ref{fig:spinfull}, where the
anomalous Hall and spin Hall effects are marked by open symbols.

The most promising path for realizing our proposal is in a cold-atoms
contexts.  A cold-atoms
realization would consist of using two species of Fermionic
gases. The first would be confined to a honeycomb lattice (and could
have one or two hyperfine states to imitate spinless or spin-half
particles). The second species would be confined to the honeycomb
lattice plane, but without being sensitive to the optical lattice that
traps the first species. This setup allows the realization of contact interactions
between the two species through a Feshbach resonance, as well as the on-site
intraspecies interactions described in Eq. (\ref{eq:int}) without significant limitations.

A realization of the scheme using electrons would employ a gated
graphene layer. Unfortunately, it suffers from significant drawbacks, since it is a long-range Coulomb interaction
which couples between the Graphene and the gate. First, when $U=0,t$ (see Fig.~\ref{fig:spinfull}-a,b),
the coupling constant $\alpha^2$ that is needed in order to induce
topological order is about $2t^2$.  Defining $d$ as the distance
between the layers, the interaction parameter is
$\alpha=\frac{e^2}{d}\frac{\pi}{k_F a}\exp(-2k_F d)$. The exponential
suppression is the result of convolving the density modulations in
the gate with the Coulomb potential as discussed in Appendix
\ref{ap:estimate}. Even choosing a gate as close as $d=6\AA$ yields
$\alpha\sim t/50$.

In addition, the large wavevector needed in the gate corresponds to a large density of
the two dimensional electron gas which is very difficult to achieve in metals or
semiconductors. Perhaps this could be mitigated by using doped graphene as
the metallic layer to induce the interactions.  At low density, the
Friedel oscillations and resulting RKKY interactions in graphene are
the result of scattering in the vicinity of the same valley point,
i.e., short wavevector, and between different valleys.  While
inter-valley scattering is suppressed due to the Klein paradox and
decay as $1/r^3$, the inter-valley scattering is allowed and is of the
right order of magnitude for our purposes \cite{Bena,TPB}.  This
suggests that the ideal gate for the realization of topological
phases (if sufficient proximity could be achieved) is another layer of graphene that is rotates by $30^o$ with
respect to the bottom one and is doped such that the inter-valley
scattering vector matches the ideal wave-vector. The challenge then
becomes to prevent the tunneling between the two rather close Graphene
planes. Note that a similar setup was considered in Refs. \cite{Macdonald-graphene,Franz-excitons,Jogelkar}

\section{Acknowledgement}
The authors like to acknowledge useful discussions with J. Alicea,
M. Franz, J. Lau, N. H. Lindner, and J. Simon. G.R. is grateful for the
generous support of the Packard Foundation and the FENA Focus
Center, one of six research centers funded under the Focus Center Research Program (FCRP), a Semiconductor
Research Corporation entity. T.P.B. and G.R are supported by the
Research Corporation Cottrell Scholars Award, and DARPA.  T.P.B. was also supported by the National Science and Engineering Council of Canada.

\appendix
\begin{widetext}
\section{Estimate of the induced interaction in the graphene layer}\label{ap:estimate}
In this section we concentrate on the electronic realization of our
proposal, and consider the details of gate-induced interaction between
two points on a graphene sheet, $r'_1$ and $r'_2$, whose distance is
$R$, taking into account screening, and the distance $d$ to the gate.  We use the known result for the interaction between these points and two points on the metallic layer $r_1$ and $r_2$ which are close to $r'_1$ and $r'_2$ but not necessarily the same.  The interaction term is therefore:
\begin{eqnarray}
e^4\int d^2 r'_1 d^2 r'_2 \int d^2r_1 d^2r_2 { n^g(r'_1) n^m(r_1) n^g(r'_2) n^m(r_2) \over |r_1-r'_2||r_2-r'_2|}
\end{eqnarray}
We would like to integrate out the metal by integrating over the positions $r_i$:
\begin{eqnarray}
\approx e^4 \int d^2 r'_1 d^2 r'_2n^g(r'_1) n^g(r'_2)\int d^2r_1 d^2r_2 {\langle n^m(r_1) n^m(r_2) \rangle \over |r_1-r'_2||r_2-r'_2|} \nonumber \\
= \int d^2 r'_1 d^2 r'_2n^g(r'_1) n^g(r'_2)V_{eff}(r'_1,r'_2)
\end{eqnarray}
The effective interaction can be calculated using the polarization of the metal:
\begin{eqnarray}
V_{eff} = e^4{1\over a^2}\int d^2r_1 d^2r_2 {\langle n^m(r_1) n^m(r_2) \rangle \over |r_1-r'_1||r_2-r'_2|} \nonumber \\
= -e^4 D(0) {1\over a^2} \int d^2\rho_1 d^2\rho_2 {\cos(2k_F(\rho_1-\rho_2)) \over (2k_F(\rho_1-\rho_2))^2}{1\over \sqrt{d^2+ (\rho_1-\rho'_1)^2}\sqrt{d^2+(\rho_2-\rho'_2)^2}}
\end{eqnarray}
where $D(0) = 1/2ta^2$ is the metallic density of states at the Fermi level  and $a \approx 2.4\AA$ is the lattice constant. We have explicitly written the distances $|r_1-r'_1|$ and $|r_2-r'_2|$ taking into account the distance between the layers $d$ and denoting the distance in each layer by $\rho$.  We rewrite the effective interaction:
\begin{eqnarray}
V_{eff} = -{e^4  \over 2 t a^4}\int \rho_1 d\rho_1 d\theta_1 \int \rho_2 d\rho_2 d\theta_2 {\cos(2k_F x) \over(2k_F x)^2 \sqrt{(d^2 + \rho_1^2)(d^2 + \rho_2^2)}}
\end{eqnarray}
where we measure $\rho_1$ from $\rho'_1$  and $\rho_2$ from $\rho'_2$ in the plane.  With this choice $x = \sqrt{R^2+\rho_1^2+\rho_2^2 -2 R(\rho_1\cos(\theta_1)-\rho_2\cos(\theta_2))-2\rho_1\rho_2\cos(\theta_1-\theta_2)}$ is the distance between $\rho_1$ and $\rho_2$.  Now, we assume that $R = |\rho'_1-\rho'_2|$ is larger than $\rho_1,\rho_2$ in the relevant part of the integral (basically assuming $|R| \ll d$) and we can simplify the expression.  We also replace $x^2$ in the denominator by $R^2$ and linearize in $r/R$ in the cosine.  This gives:
\begin{eqnarray}
V_{eff} &\approx& -{e^4  \over 2 t a^4 (2 k_F R)^2}\int \rho_1 d\rho_1 d\theta_1 \int \rho_2 d\rho_2 d\theta_2 {\cos(2k_F (R - \rho_1\cos(\theta_1) + \rho_2\cos(\theta_2))) \over  \sqrt{(d^2 + \rho_1^2)(d^2 + \rho_2^2)}} \nonumber \\
&=&-2\pi{e^4  \over 2 t a^4 (2 k_F R)^2}\int \rho_1 d\rho_1 \int \rho_2 d\rho_2 d\theta_2  {J_0(2k_F \rho_1)\cos(2k_F (R + \rho_2\cos(\theta_2))) \over  \sqrt{(d^2 + \rho_1^2)(d^2 + \rho_2^2)}} \nonumber \\
&=& -(2\pi)^2{e^4  \over 2 t a^4 (2 k_F R)^2}\int \rho_1 d\rho_1 \int \rho_2 d\rho_2  {J_0(2k_F \rho_1)J_0(2k_F \rho_2)\cos(2k_F R) \over  \sqrt{(d^2 + \rho_1^2)(d^2 + \rho_2^2)}} \nonumber \\
&=& -(2\pi)^2{e^4 \cos(2k_F R) \over 2 t a^4 (2 k_F R)^2}\left( \int \rho d\rho  {J_0(2k_F \rho) \over  \sqrt{(d^2 + \rho^2)}}\right)^2 \nonumber \\
 &=&  -(2\pi)^2{e^4 \cos(2k_F R) \over 2 t a^4 (2 k_F R)^2}\left( {e^{-2k_F d} \over 2 k_F}\right)^2 \nonumber \\
 &=&  -\pi^2{e^4 \cos(2k_F R) \over 8 t a^4 k_F^4 R^2} e^{-4k_F d}
\end{eqnarray}
where we have used $\int d\theta \cos(a + b\cos(\theta)) = 2\pi J_0(b)\cos(a)$ in the angular integrals.
Indeed, this expression is suppressed by ${(2\pi)^2 \over (2k_F a)^2}\exp(-4k_F d)$ with respect to the bare coupling square.
\end{widetext}

\bibliographystyle{apsrev}

\end{document}